\documentclass[article,twocolumn,prb,showpacs,superscriptaddress]{revtex4-2}

\usepackage{graphicx}
\usepackage{amssymb}
\usepackage{bm}
\usepackage{amsmath}
\usepackage{xfrac}
\usepackage{color}
\usepackage{titlesec}
\renewcommand{\figurename}{Figure} 
\makeatletter	        
\def\fnum@figure{\textbf{\figurename~\thefigure}}
\makeatother
\usepackage[normalem]{ulem}
\usepackage[utf8]{inputenc}
\usepackage[T1]{fontenc}
\usepackage[marginal]{footmisc}

\usepackage{latexsym}

\usepackage[colorlinks=True,
                 linkcolor=black,
	        citecolor=blue,
	        urlcolor=blue]{hyperref}

\makeatletter	        
 \def\section{%
  \@startsection{section}{1}{\z@}{0.8cm plus1ex minus.2ex}{0.2cm}%
  {%
   \small\sffamily\bfseries\selectfont
   \raggedright
   \parindent\z@
  }%
 }%
  \def\subsection{%
  \@startsection{subsection}{2}{\z@}{0.8cm plus1ex minus.2ex}{0.2cm}%
  {%
   \small\sffamily\bfseries
   \raggedright
   \parindent\z@
  }%
 }%
\makeatother

\newcommand{\comment}{\textcolor{black}}

\usepackage{color}

\newcommand{\MOE}{MOE Key Laboratory for Nonequilibrium Synthesis and Modulation of Condensed Matter, School of Physics, Xi’an Jiaotong University, Xi’an 710049, China}
\newcommand{\NJ}{Key Laboratory of Flexible Electronics and Institute of Advanced Materials, Nanjing Tech University, Nanjing 211816, China}
\newcommand{\QD}{State Key Lab of Metastable Materials Science and Technology, Yanshan University, Qinhuangdao 066004, China}

\newcommand{\Yanshan}{Key Laboratory for Microstructure Material Physics of Hebei Province, Yanshan University, Qinhuangdao 066004, China}

\newcommand{\ME}{State Key Laboratory for Manufacturing Systems Engineering, Xi’an Jiaotong University, Xi’an,710049, China}

\makeatletter
\g@addto@macro\bfseries{\boldmath}
\makeatother

\begin{document}
\title{Bias Voltage Driven Tunneling Magnetoresistance Polarity Reversal in 2D Stripy Antiferromagnet CrOCl}

\author{Lihao Zhang}
\author{Xiaoyu Wang}
\affiliation{\MOE}
\author{Qi Li}
\affiliation{\NJ}
\author{Haibo Xie}
\affiliation{\MOE}
\author{Liangliang Zhang}
\affiliation{\ME}
\author{Lei Zhang}
\affiliation{\MOE}
\author{Jie Pan}
\email{jiepan@xjtu.edu.cn}
\affiliation{\MOE}

\author{Yingchun Cheng}
\email{iamyccheng@ysu.edu.cn}
\affiliation{\Yanshan}
\affiliation{\QD}
\author{Zhe Wang}
\email{zhe.wang@xjtu.edu.cn}
\affiliation{\MOE}



\begin{abstract}

Atomically thin materials with coupled magnetic and electric polarization are critical for developing energy-efficient and high-density spintronic devices, yet they remain scarce due to often conflicting requirements of stabilizing both magnetic and electric orders. The recent discovery of the magnetoelectric effect in the 2D stripy antiferromagnet CrOCl highlights this semiconductor as a promising platform to explore electric field effects on magnetoresistance. In this study, we systematically investigate the magnetoresistance in tunneling junctions of bilayer and monolayer CrOCl. We observe that the transition from antiferromagnetic to ferrimagnetic phases in both cases induces a positive magnetoresistance at low bias voltages, which reverses to a negative value at higher bias voltages. This polarity reversal is attributed to the additional electric dipoles present in the antiferromagnetic state, as supported by our theoretical calculations. These findings suggest a pathway for the electric control of spintronic devices and underscore the potential of 2D magnets like CrOCl in advancing energy-efficient spintronic applications.   
 
\end{abstract}

\maketitle


The tunneling magnetoresistance (TMR) effect is foundational to spintronic devices, which are crucial for information storage and logic technologies~\cite{2fert2008nobel,5dietl2014dilute,13gould2004tunneling,14heiliger2007tunneling,8inoue1996theory,4jungwirth2014spin,7jungwirth2016antiferromagnetic,6manchon2019current,10mathon2001theory,3sato2010first,1vzutic2004spintronics,15wolf2001spintronics}. Controlling TMR via bias voltage is vital for achieving low power consumption. TMR polarity reversals have been observed in various magnetic tunneling junctions comprising ferromagnetic metal/insulator/ferromagnetic metal structures~\cite{16morosov2014ferromagnetic,17min2022tunable,19wang2012electric,23tiusan2004interfacial,24piquemal2018insulator,27de1999inverse,29zhu2023large,30zhu2024voltage,31jin2023room,32pan2023room}, where the bias voltage enables high-energy electrons to participate in the tunneling process, resulting in a reversal of spin polarization. Typically, semiconductors are more easily controlled by electric fields due to the absence of electron screening effects. However, TMR polarity reversal driven by bias voltage has rarely yet been reported in tunneling structures consisting of magnetic semiconductors. This is largely due to the rarity of materials with coupled magnetic and electric polarization in such systems.

CrOCl, an air stable 2D stripy antiferromagnetic semiconductor, has recently attracted considerable attention due to its intriguing magnetic and electronic properties~\cite{34zhang2019magnetism,44gu2022magnetic,48gu2023multi,37zhang2023spin,35qing2020magnetism,36angelkort2009observation,38schaller2023pressure,40zhang2022tuning,42yang2023unconventional,33guo2024van}. It transforms into an antiferromagnetic (AFM) ground state ↑↑↓↓with a periodicity of 4b below Neel temperature of around 14 K, accompanied by a structural transition from orthorhombic space group Pmmn to monoclinic space group P2$_1$/m~\cite{25_49,26_50,27_51,30_54}. As schematic presented in \comment{the inset of Fig.1b}, upon applying a strong enough magnetic field along c-axis, it undergoes a transition from AFM to a ferrimagnetic (FiM) state ↑↑↑↓↓~\cite{35qing2020magnetism,36angelkort2009observation,37zhang2023spin,44gu2022magnetic,48gu2023multi,30_54}, with a concurrent reversion of the crystal structure back to the orthorhombic phase~\cite{48gu2023multi}. This magnetoelastic coupling has been observed to persist down to thin exfoliated flakes~\cite{44gu2022magnetic}, as evidenced by Raman measurements. Notably, recent pioneering studies have also revealed that CrOCl exhibits intrinsic magnetoelectric properties~\cite{48gu2023multi}, with additional polarization observed in both bulk and atomically thin flakes, making it an excellent platform for exploring bias voltage effects on TMR. By employing atomically thin CrOCl as a tunneling barrier with graphene electrodes, we systematically investigate TMR at various bias voltages. Our findings indicate that magnetic transitions significantly influence tunneling resistance, with TMR being positive at low bias voltages and reversing to negative at higher voltages. Remarkably, this TMR polarity reversal persists even down to the monolayer limit.


\begin{figure*}
\centering
\includegraphics[width =1\linewidth]{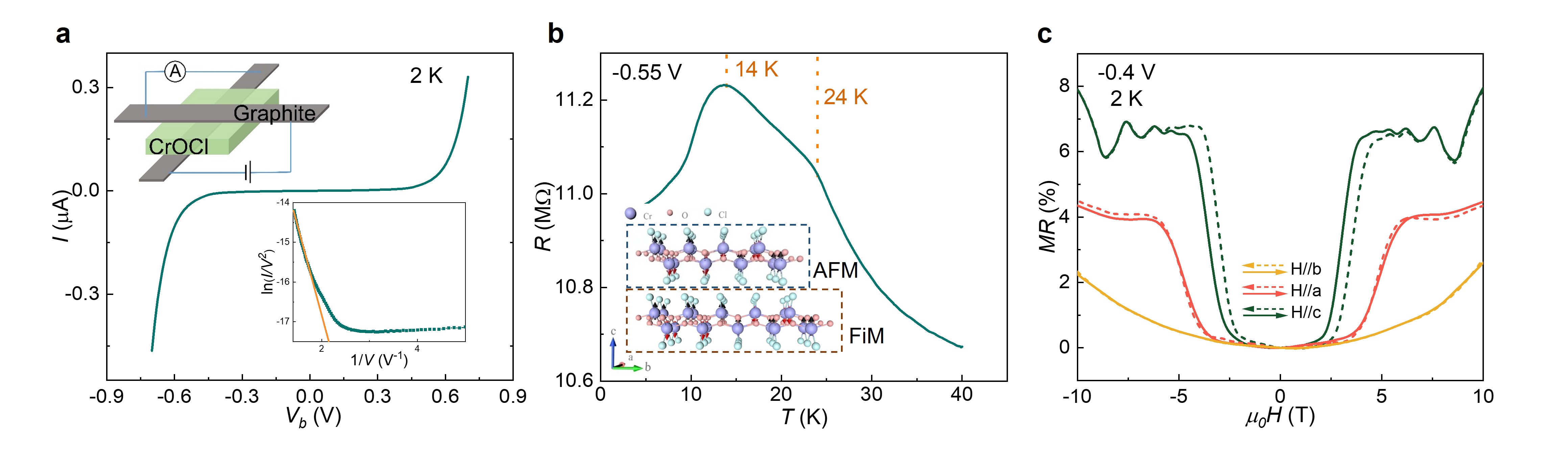}
\caption{\textbf{Magnetic phase transition detected in bilayer CrOCl} \textbf{a}. Bias voltage dependence of current for tunnleing junction of bilayer-CrOCl at 2 K. The up inset show the schematic of tunneling junctions with graphite serving as electrodes. The down inset show the plot of ln(I/V²) as a function of 1/V. \textbf{b}. Temperature dependence of tunneling resistance at zero magnetic field. \comment{The inset show crystal and magnetic structure of CrOCl, depicting its stripy antiferromagnetic behavior below the Néel temperature with the easy axis along the c-axis (up figure). It transforms into ferrimagnetic phase upon a strong magnetic field applied along c-axis (down figure).}\textbf{c}. TMR at 2 K with magnetic field applied along different crystal axis of CrOCl.} 
\end{figure*}


Single crystals of CrOCl were grown using the chemical vapor transport method following established protocols~\cite{44gu2022magnetic,26_50}. \comment{The magnetization of grown crystal is characterized with a SQUID magnetometer and the results are shown in the Supplementary Materials. To fabricate the tunneling junctions as shown in the upper inset of Fig. 1d, atomically thin flakes of hBN, graphene and CrOCl were firstly obtained via mechanical exfoliation from bulk crystals in a nitrogen-filled glove box to minimize contaminations. van der Waals heterostructures were then assembled using a "pick-up" method with PDMS/PC stamp. Finally, contacts of a 10 nm/50 nm Cr/Au were deposited onto the graphene electrodes through a three-step process, including electron beam lithography for precise patterning, reactive ion etching for etching hBN, and electron beam evaporation for metal deposition. The devices were then loaded into a cryostat from Cryogenic Limited, which provide low temperature and high magnetic field measurement environment. Electrical transport measurements were performed using a constant voltage method with a Keithley 2400 and SR830, and frequency of 17.377 was used for AC measurement. The resistance of few-layer graphene eletrodes is typically smaller than 1 kilohm, several orders smaller than the resistance of tunneling junctions.}

\begin{figure*}
\includegraphics[width =1\linewidth]{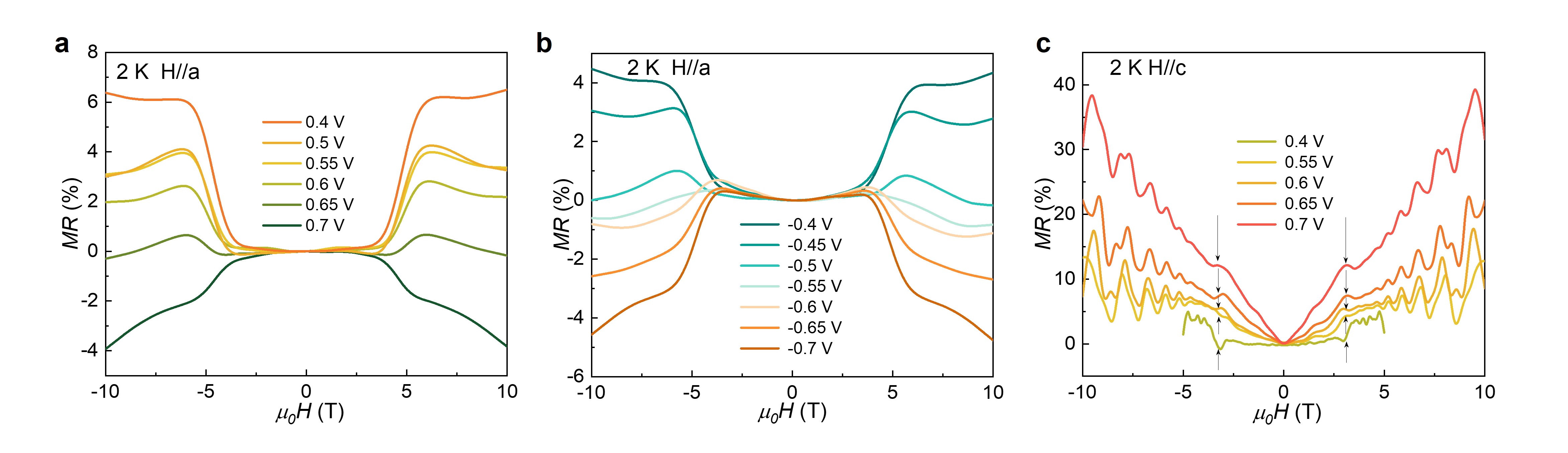}
\caption{\textbf{Bias voltage induced TMR polarity reversal in bilayer CrOCl.}   \textbf{a} and \textbf{b}. TMR measured with different positive and negative bias voltage respectively, with magnetic field applied along a-axis. \textbf{c} TMR measured with different positive voltage and magnetic field applied along c-axis.  \comment{The data of negative bias voltage is shown in supplementary materials.}} 
\end{figure*}

The magnetic phase transition in atomically thin CrOCl flakes can be detected through electronic tunneling. The voltage dependence of current at 2K, shown in Fig. 1a, is non-linear, as expected for a typical tunneling junction. Plotting ln(I/V²) as a function of 1/V reveals a linear relationship, indicating Fowler-Nordheim tunneling as observed in different 2D magnetic semiconductor vertical junctions~\cite{31_55,32_56,33_57}. Fig. 1b shows the temperature dependence of resistance at zero field. Resistance slightly increases with temperature, with a notable change in the rate of increases around 24 K, close to the ~27 K detected in bulk magnetization measurements. Below 14 K, resistance stops increasing and begins to decreases, consistent with CrOCl's antiferromagnetic transition temperature. Applying a magnetic field along the c-axis causes a sharp increases in resistance around 3 Tesla, consistent with the magnetic field required for the transition from AFM to FiM observed in bulk magnetization measurements. \comment{The MR oscillations at high magnetic field is due to changes in the density of states in graphene electrodes under magnetic fields.}



The sharp edges of exfoliated flakes enable investigation of phase transitions with magnetic fields applied along different crystallographic axes. Fig. \comment{1c} presents TMR results at 2K, here solid lines indicate magnetic field sweeps from negative to positive values, while dashed lines represent reverse sweep. A sharp increases in tunneling resistance occurs when the magnetic field is applied along the a-axis, indicating a phase transition from AFM to FiM, with the transition field slightly higher than that for the c-axis. In contrast, tunneling resistance changes smoothly when the magnetic field is applied along the b-axis. These observations align with previous studies, which identified the a-c plane as the easy plane in the FiM state\cite{37zhang2023spin}. 


After confirming that magnetic phase transitions influence tunneling resistance, we explored the effect of bias voltage on TMR. Fig. 2\comment{a} presents TMR results with the magnetic field parallel to the a-axis. TMR is positive at low bias voltages, meaning resistance is lower in the AFM state than in the FiM state. As bias voltage increases, TMR amplitude decreases, eventually becoming negative at 0.7 V. The TMR polarity reversal is also observed for negative bias voltages, as shown in Fig. 2\comment{b}, where the resistance of the AFM state becomes higher than that of the FiM state when the voltage is below -0.55 V.

The bias voltage-induced TMR polarity reversal is also observed when the magnetic field is applied along the c-axis, \comment{as shown in Fig.2c}. Due to changes in the density of states in graphene electrodes and even the development of Landau levels under magnetic fields, TMR exhibits a substantial background. However, focusing on the region around 3 Tesla, where the transition from AFM to FiM occurs (indicated by arrows), the TMR polarity reversal driven by bias voltage is still observable. No measurements with higher bias voltages were conducted for the b-axis field, as no magnetic phase transition is observed in this configuration. All behavior described above are repeated in another bilayer device, whose data is presented in the supplementary materials.

\begin{figure}
\centering
\includegraphics[width =1\linewidth]{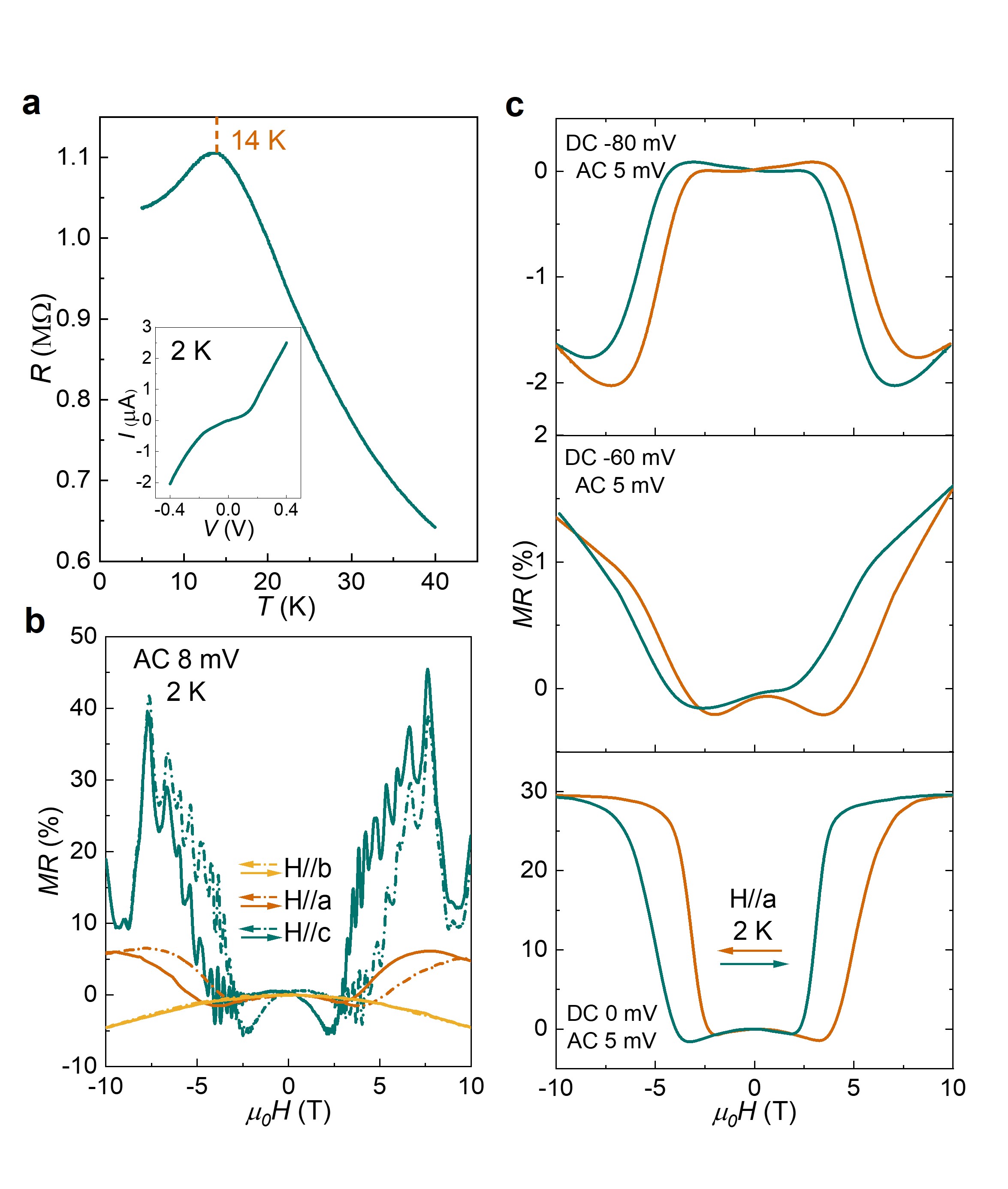}
\caption{\textbf{Bias voltage induced TMR polarity reversal in mono-layer CrOCl.} \textbf{a}. Temperature dependence of tunneling resistance at zero magnetic field, measured with constant AC bias voltage of 8 mV. The inset show the $IV$ curve at 2 K. \textbf{b}. TMR at 2 K with magnetic field applied along different crystal axis. \textbf{c}. Magnetic field dependence of differential resistance measured with different DC bias voltage, the AC excitation voltage is 5 mV. } 
\end{figure}

We further explored this effect in atomically thin CrOCl, specifically the monolayer case. The inset of Fig. 3a shows the IV curve of monolayer CrOCl device, while the main figure illustrates the temperature dependence of the tunneling resistance. The behavior is similar to the bilayer device shown in Fig. 1a: resistance increases as the temperature lowers and begins to decreases upon entering the AFM state. The Néel temperature obtained from this measurement is around 14 K, identical to that of the bilayer and bulk samples, indicating weak effect of interlayer coupling on intralayer magnetic order. Notably, the monolayer device does not display the characteristic feature at ~27 K, where a magnetic transition from the paramagnetic state to an incommensurate spin density wave is observed in bulk and bilayer samples. 

Figure 3b presents TMR data for monolayer CrOCl with the magnetic field applied along the three principal crystallographic axes. For the field along the c-axis (the easy axis), the resistance begins to increases around 2.2 Tesla, consistent with the bilayer sample, which shows a marked increases between 2.2 and 4.0 Tesla. The oscillations in TMR, attributed to the graphene electrodes, are more pronounced in the monolayer device, which also exhibits greater hysteresis during magnetic field sweeps. With the magnetic field applied along the a-axis, the resistance increases around 3.6 Tesla, again aligning closely with the bilayer results but showing more significant hysteresis in the monolayer. When the magnetic field is applied along the b-axis, the TMR shows a slight decrease across the entire range of the magnetic field with no observable hysteresis. To mitigate the influence of graphene electrode density of states oscillations, we focused on data with the magnetic field parallel to the a-axis. As shown in Fig. \comment{Supplementary Fig.4a}, TMR due to magnetic phase transition decreases in amplitude with increasing temperature and disappears above 14 K, close to the Néel temperature identified in \comment{bilayer samples}. These measurements confirm that magnetic fields along the c-axis and a-axis induce a phase transition from AFM to FiM states in the monolayer device, similar to multilayer CrOCl.

After confirming the magnetic field-induced phase transition in the monolayer limit, we examined TMR behavior at various bias voltages. \comment{Supplementary Fig.4a} shows the TMR measured at different DC bias voltages. Although the amplitude of TMR decreases with increasing bias voltage, no clear positive TMR signal is observed up to the highest voltage of ±0.4 V. We notice that the hysteresis vanishes for magnetic sweeps conducted with high DC bias voltages, contrasting with the pronounced hysteresis seen under low AC voltage excitations. To avoid Joule heating due to excessive tunneling current and potential device damage, we opted for differential resistance measurements with combined DC and small AC bias voltages. The differential resistance reveals local conductance changes at specific energy levels, isolating the effects from zero to the applied voltage. Figure 3c shows results with an AC voltage of 5 mV and DC bias voltages of 0 V, -60 mV, and -80 mV. TMR is positive at 0 and -60 mV bias but reverses to negative at -80 mV. On the positive voltage side, TMR remains near zero with increased noise at 80 mV. Despite these limitations, these data demonstrate the TMR polarity reversal in monolayer CrOCl. 
 
\begin{figure}[t]
\includegraphics[width =1\linewidth]{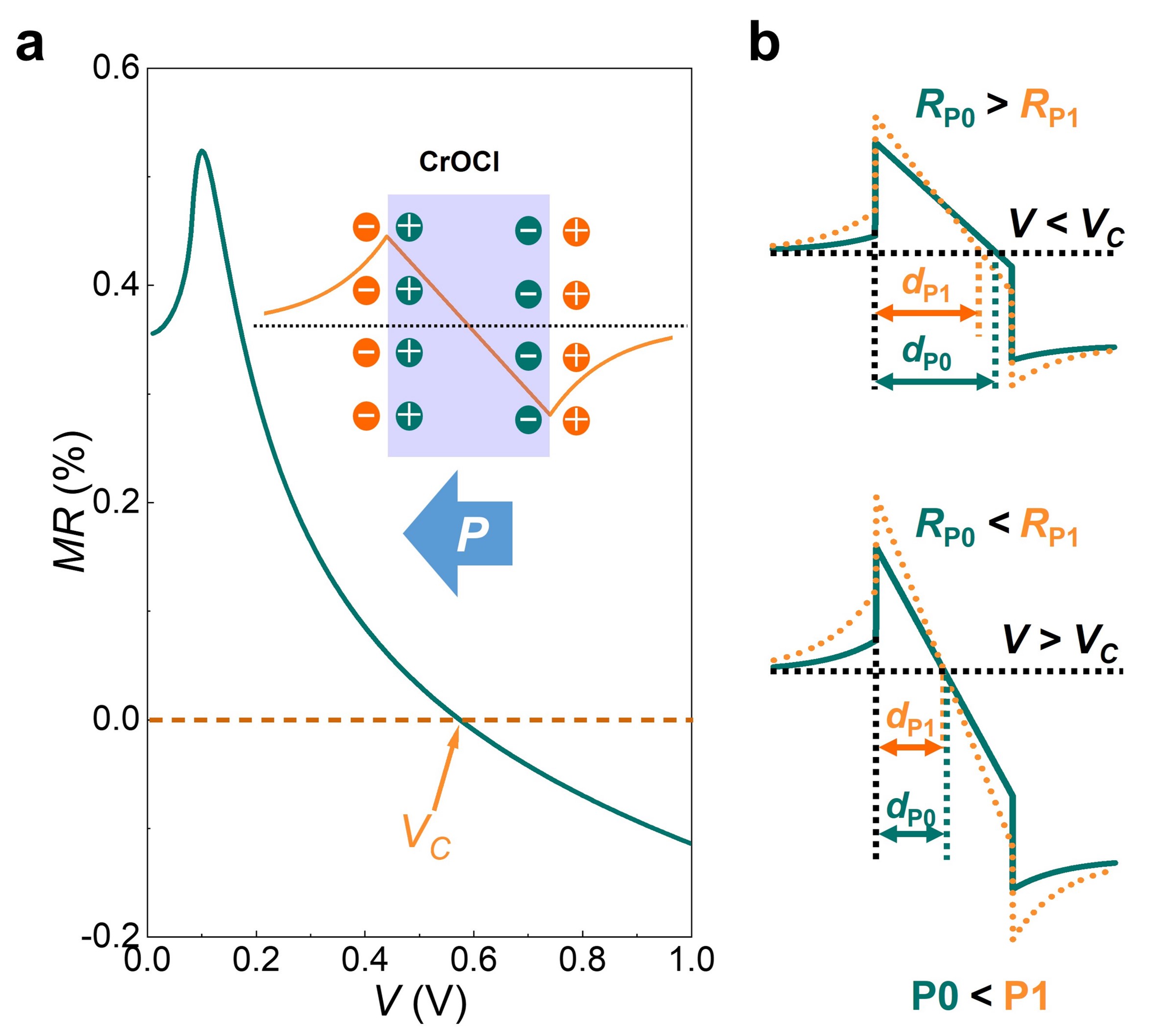}
\caption{\textbf{The theoretical model of TMR polarity reversal in CrOCl. } \textbf{a}. Bias voltage dependence of calculated TMR between antiferromagnetic and ferrimagnetic phase. The inset is schematic picture showing the influence of electric dipole on the potential profile of tunneling junction. \textbf{b}. Schematic pictures of potential profile for small electric dipole (green line) and large electric dipole (yellow line).} 
\end{figure}

Typically, TMR polarity does not change with bias voltage in tunneling junctions involving 2D magnetic semiconductors like CrX$_3$ (X = I, Br, Cl)~\cite{33_57,58yao2023multiple,38_59}, CrSBr~\cite{60telford2020layered}, and MnPS$_3$~\cite{61}. The unique observations in CrOCl can be attributed to its magnetoelectric properties, which are absent in other 2D magnetic semiconductors mentioned above. Recent measurements of dielectric properties and calculations indicate a decrease in relative permittivity from 3.07 to 3.06 as CrOCl transitions from AFM to FiM~\cite{48gu2023multi}. This occurs because the stripy AFM phase, with monoclinic distortion, can be viewed as an antiferroelectric order, allowing an external electric field to induce additional electric dipoles. Conversely, in the FiM phase, the structure relaxes to an orthorhombic form, rendering it less responsive to external electric fields~\cite{48gu2023multi}. The induced electric dipole will lead to charge carrier accumulation on CrOCl surface. Given that graphene's semimetallic nature and small density of states hinder effective charge screening, the surface charge carriers of CrOCl influence the potential profile of the tunneling junctions, as depicted in the inset of Fig. 4a. Using measured relative permittivity values from reference~\cite{48gu2023multi} and calculations based on the WKB approximation (details in the supplementary materials), we modeled the tunneling conductance. To align with the observation of positive TMR at zero bias, we assumed a slightly higher tunneling barrier for the FiM phase compared to the AFM phase. Figure 4a shows the calculated bias voltage dependence of TMR, revealing a polarity reversal when the voltage surpasses a critical threshold Vc. Supplementary materials further illustrate the differential resistance results, showing a significantly reduced Vc compared to resistance results, which qualitatively explains the observations in monolayer devices. 

Fig. 4b schematically explains the mechanism behind TMR polarity reversal. The potential profile influenced by the electric dipole impacts the tunneling process in two ways: it narrows the effective tunneling barrier width and increases the barrier height. At low bias voltages, the narrowing effect dominates, leading to lower resistance with a larger electric dipole. At higher electric fields, the effective barrier width is mainly determined by the electric field, so the increased barrier height due to the electric dipole becomes the dominant factor, resulting in lower resistance with a smaller dipole. Although many factors are not considered in our calculations, this simplified model captures the key findings of our work.                

In summary, we have systematically investigated the resistance of CrOCl tunneling junctions as a function of magnetic field amplitude and direction, temperature, and bias voltage. Our results reveal that the magnetic phase transition from AFM to FiM states significantly alters the tunneling resistance, with the a-c plane identified as the easy plane in the FiM state. Notably, the TMR exhibits a positive polarity at low bias voltages and reverses to negative at higher bias voltages, a phenomenon that persists down to the monolayer limit. Supported by our calculations, this polarity reversal is attributed to changes in the relative permittivity between the AFM and FiM states, driven by CrOCl's intrinsic magnetoelectric properties. These findings establish CrOCl as a model system for probing magnetoelectric coupling effects at the atomic scale and highlight its potential for developing energy-efficient spintronic devices that leverage the interplay between magnetic and electric orders.       

\section*{Supplementary Material}
\comment{Magnetization data for CrOCl bulk crystals, additional data for devices discussed in the main text, electronic transport data for a second bilayer device, and tunneling resistance calculations are included in the supplementary materials.} 



\section*{Acknowledgement}
This work is financially supported by National Natural Science Foundation of China (Grants no. 12374121, 12304232, 12274090 and 62274087), Shaanxi Fundamental Science Research Project for Mathematics and Physics (22JSY026,23JSQ011).
\bibliographystyle{mynaturemag}
\bibliography{biblio.bib}

\begin{thebibliography}{10}
\expandafter\ifx\csname url\endcsname\relax
  \def\url#1{\texttt{#1}}\fi
\expandafter\ifx\csname urlprefix\endcsname\relax\def\urlprefix{URL }\fi
\providecommand{\bibinfo}[2]{#2}
\providecommand{\eprint}[2][]{\url{#2}}

\bibitem{2fert2008nobel}
\bibinfo{author}{Fert, A.}
\newblock \bibinfo{title}{Nobel lecture: Origin, development, and future of spintronics}.
\newblock \emph{\bibinfo{journal}{Reviews of Modern Physics}} \textbf{\bibinfo{volume}{80}}\textbf{, }\bibinfo{pages}{1517--1530} (\bibinfo{year}{2008}).

\bibitem{5dietl2014dilute}
\bibinfo{author}{Dietl, T.} \& \bibinfo{author}{Ohno, H.}
\newblock \bibinfo{title}{Dilute ferromagnetic semiconductors: Physics and spintronic structures}.
\newblock \emph{\bibinfo{journal}{Reviews of Modern Physics}} \textbf{\bibinfo{volume}{86}}\textbf{, }\bibinfo{pages}{187--251} (\bibinfo{year}{2014}).

\bibitem{13gould2004tunneling}
\bibinfo{author}{Gould, C.}, \bibinfo{author}{R\"uster, C.}, \bibinfo{author}{Jungwirth, T.} \emph{et~al.}
\newblock \bibinfo{title}{Tunneling anisotropic magnetoresistance: A spin-valve-like tunnel magnetoresistance using a single magnetic layer}.
\newblock \emph{\bibinfo{journal}{Physical Review Letters}} \textbf{\bibinfo{volume}{93}}\textbf{, }\bibinfo{pages}{117203} (\bibinfo{year}{2004}).

\bibitem{14heiliger2007tunneling}
\bibinfo{author}{Heiliger, C.}, \bibinfo{author}{Gradhand, M.}, \bibinfo{author}{Zahn, P.} \emph{et~al.}
\newblock \bibinfo{title}{Tunneling magnetoresistance on the subnanometer scale}.
\newblock \emph{\bibinfo{journal}{Physical Review Letters}} \textbf{\bibinfo{volume}{99}}\textbf{, }\bibinfo{pages}{066804} (\bibinfo{year}{2007}).

\bibitem{8inoue1996theory}
\bibinfo{author}{Inoue, J.} \& \bibinfo{author}{Maekawa, S.}
\newblock \bibinfo{title}{Theory of tunneling magnetoresistance in granular magnetic films}.
\newblock \emph{\bibinfo{journal}{Physical Review B}} \textbf{\bibinfo{volume}{53}}\textbf{, }\bibinfo{pages}{R11927} (\bibinfo{year}{1996}).

\bibitem{4jungwirth2014spin}
\bibinfo{author}{Jungwirth, T.}, \bibinfo{author}{Wunderlich, J.}, \bibinfo{author}{Nov{\'a}k, V.} \emph{et~al.}
\newblock \bibinfo{title}{Spin-dependent phenomena and device concepts explored in {(Ga, Mn) As}}.
\newblock \emph{\bibinfo{journal}{Reviews of Modern Physics}} \textbf{\bibinfo{volume}{86}}\textbf{, }\bibinfo{pages}{855--896} (\bibinfo{year}{2014}).

\bibitem{7jungwirth2016antiferromagnetic}
\bibinfo{author}{Jungwirth, T.}, \bibinfo{author}{Marti, X.}, \bibinfo{author}{Wadley, P.} \emph{et~al.}
\newblock \bibinfo{title}{Antiferromagnetic spintronics}.
\newblock \emph{\bibinfo{journal}{Nature Nanotechnology}} \textbf{\bibinfo{volume}{11}}\textbf{, }\bibinfo{pages}{231--241} (\bibinfo{year}{2016}).

\bibitem{6manchon2019current}
\bibinfo{author}{Manchon, A.}, \bibinfo{author}{{\v{Z}}elezn{\`y}, J.}, \bibinfo{author}{Miron, I.~M.} \emph{et~al.}
\newblock \bibinfo{title}{Current-induced spin-orbit torques in ferromagnetic and antiferromagnetic systems}.
\newblock \emph{\bibinfo{journal}{Reviews of Modern Physics}} \textbf{\bibinfo{volume}{91}}\textbf{, }\bibinfo{pages}{035004} (\bibinfo{year}{2019}).

\bibitem{10mathon2001theory}
\bibinfo{author}{Mathon, J.} \& \bibinfo{author}{Umerski, A.}
\newblock \bibinfo{title}{Theory of tunneling magnetoresistance of an epitaxial {Fe/MgO/Fe (001)} junction}.
\newblock \emph{\bibinfo{journal}{Physical Review B}} \textbf{\bibinfo{volume}{63}}\textbf{, }\bibinfo{pages}{220403} (\bibinfo{year}{2001}).

\bibitem{3sato2010first}
\bibinfo{author}{Sato, K.}, \bibinfo{author}{Bergqvist, L.}, \bibinfo{author}{Kudrnovsk{\`y}, J.} \emph{et~al.}
\newblock \bibinfo{title}{First-principles theory of dilute magnetic semiconductors}.
\newblock \emph{\bibinfo{journal}{Reviews of Modern Physics}} \textbf{\bibinfo{volume}{82}}\textbf{, }\bibinfo{pages}{1633--1690} (\bibinfo{year}{2010}).

\bibitem{1vzutic2004spintronics}
\bibinfo{author}{{\v{Z}}uti{\'c}, I.}, \bibinfo{author}{Fabian, J.} \& \bibinfo{author}{Sarma, S.~D.}
\newblock \bibinfo{title}{Spintronics: Fundamentals and applications}.
\newblock \emph{\bibinfo{journal}{Reviews of Modern Physics}} \textbf{\bibinfo{volume}{76}}\textbf{, }\bibinfo{pages}{323} (\bibinfo{year}{2004}).

\bibitem{15wolf2001spintronics}
\bibinfo{author}{Wolf, S.}, \bibinfo{author}{Awschalom, D.}, \bibinfo{author}{Buhrman, R.} \emph{et~al.}
\newblock \bibinfo{title}{Spintronics: a spin-based electronics vision for the future}.
\newblock \emph{\bibinfo{journal}{Science}} \textbf{\bibinfo{volume}{294}}\textbf{, }\bibinfo{pages}{1488--1495} (\bibinfo{year}{2001}).

\bibitem{16morosov2014ferromagnetic}
\bibinfo{author}{Morosov, A.}
\newblock \bibinfo{title}{Ferromagnetic magnetization switching by an electric field: A review}.
\newblock \emph{\bibinfo{journal}{Physics of the Solid State}} \textbf{\bibinfo{volume}{56}}\textbf{, }\bibinfo{pages}{865--872} (\bibinfo{year}{2014}).

\bibitem{17min2022tunable}
\bibinfo{author}{Min, K.-H.}, \bibinfo{author}{Lee, D.~H.}, \bibinfo{author}{Choi, S.-J.} \emph{et~al.}
\newblock \bibinfo{title}{Tunable spin injection and detection across a van der waals interface}.
\newblock \emph{\bibinfo{journal}{Nature Materials}} \textbf{\bibinfo{volume}{21}}\textbf{, }\bibinfo{pages}{1144--1149} (\bibinfo{year}{2022}).

\bibitem{19wang2012electric}
\bibinfo{author}{Wang, W.-G.}, \bibinfo{author}{Li, M.}, \bibinfo{author}{Hageman, S.} \emph{et~al.}
\newblock \bibinfo{title}{Electric-field-assisted switching in magnetic tunnel junctions}.
\newblock \emph{\bibinfo{journal}{Nature Materials}} \textbf{\bibinfo{volume}{11}}\textbf{, }\bibinfo{pages}{64--68} (\bibinfo{year}{2012}).

\bibitem{23tiusan2004interfacial}
\bibinfo{author}{Tiusan, C.}, \bibinfo{author}{Faure-Vincent, J.}, \bibinfo{author}{Bellouard, C.} \emph{et~al.}
\newblock \bibinfo{title}{Interfacial resonance state probed by spin-polarized tunneling in epitaxial $\mathrm{F}\mathrm{e}/\mathrm{M}\mathrm{g}\mathrm{O}/\mathrm{F}\mathrm{e}$ tunnel junctions}.
\newblock \emph{\bibinfo{journal}{Physical Review Letters}} \textbf{\bibinfo{volume}{93}}\textbf{, }\bibinfo{pages}{106602} (\bibinfo{year}{2004}).

\bibitem{24piquemal2018insulator}
\bibinfo{author}{Piquemal-Banci, M.}, \bibinfo{author}{Galceran, R.}, \bibinfo{author}{Godel, F.} \emph{et~al.}
\newblock \bibinfo{title}{Insulator-to-metallic spin-filtering in 2d-magnetic tunnel junctions based on hexagonal boron nitride}.
\newblock \emph{\bibinfo{journal}{ACS Nano}} \textbf{\bibinfo{volume}{12}}\textbf{, }\bibinfo{pages}{4712--4718} (\bibinfo{year}{2018}).

\bibitem{27de1999inverse}
\bibinfo{author}{De~Teresa, J.}, \bibinfo{author}{Barth{\'e}l{\'e}my, A.}, \bibinfo{author}{Fert, A.} \emph{et~al.}
\newblock \bibinfo{title}{{Inverse tunnel magnetoresistance in Co/SrTiO$_3$/La$_{0.7}$Sr$_{0.3}$MnO$_3$: new ideas on spin-polarized tunneling}}.
\newblock \emph{\bibinfo{journal}{Physical Review Letters}} \textbf{\bibinfo{volume}{82}}\textbf{, }\bibinfo{pages}{4288} (\bibinfo{year}{1999}).

\bibitem{29zhu2023large}
\bibinfo{author}{Zhu, W.}, \bibinfo{author}{Zhu, Y.}, \bibinfo{author}{Zhou, T.} \emph{et~al.}
\newblock \bibinfo{title}{Large and tunable magnetoresistance in van der waals ferromagnet/semiconductor junctions}.
\newblock \emph{\bibinfo{journal}{Nature Communications}} \textbf{\bibinfo{volume}{14}}\textbf{, }\bibinfo{pages}{5371} (\bibinfo{year}{2023}).

\bibitem{30zhu2024voltage}
\bibinfo{author}{Zhu, S.}, \bibinfo{author}{Lin, H.}, \bibinfo{author}{Zhu, W.} \emph{et~al.}
\newblock \bibinfo{title}{{Voltage tunable sign inversion of magnetoresistance in van der Waals Fe$_3$GeTe$_2$/MoSe$_2$/Fe$_3$GeTe$_2$ tunnel junctions}}.
\newblock \emph{\bibinfo{journal}{Applied Physics Letters}} \textbf{\bibinfo{volume}{124}} (\bibinfo{year}{2024}).

\bibitem{31jin2023room}
\bibinfo{author}{Jin, W.}, \bibinfo{author}{Zhang, G.}, \bibinfo{author}{Wu, H.} \emph{et~al.}
\newblock \bibinfo{title}{Room-temperature and tunable tunneling magnetoresistance in {Fe$_{3}$GaTe$_{2}$}-based 2d van der waals heterojunctions}.
\newblock \emph{\bibinfo{journal}{ACS Applied Materials \& Interfaces}} \textbf{\bibinfo{volume}{15}}\textbf{, }\bibinfo{pages}{36519--36526} (\bibinfo{year}{2023}).

\bibitem{32pan2023room}
\bibinfo{author}{Pan, H.}, \bibinfo{author}{Singh, A.~K.}, \bibinfo{author}{Zhang, C.} \emph{et~al.}
\newblock \bibinfo{title}{Room-temperature tunable tunneling magnetoresistance in {Fe$_3$GaTe$_2$/WSe$_2$/Fe$_3$GaTe$_2$} van der waals heterostructures}.
\newblock \emph{\bibinfo{journal}{InfoMat}} \bibinfo{pages}{e12504} (\bibinfo{year}{2023}).

\bibitem{34zhang2019magnetism}
\bibinfo{author}{Zhang, T.}, \bibinfo{author}{Wang, Y.}, \bibinfo{author}{Li, H.} \emph{et~al.}
\newblock \bibinfo{title}{Magnetism and optical anisotropy in van der waals antiferromagnetic insulator {CrOCl}}.
\newblock \emph{\bibinfo{journal}{ACS Nano}} \textbf{\bibinfo{volume}{13}}\textbf{, }\bibinfo{pages}{11353--11362} (\bibinfo{year}{2019}).

\bibitem{44gu2022magnetic}
\bibinfo{author}{Gu, P.}, \bibinfo{author}{Sun, Y.}, \bibinfo{author}{Wang, C.} \emph{et~al.}
\newblock \bibinfo{title}{Magnetic phase transitions and magnetoelastic coupling in a two-dimensional stripy antiferromagnet}.
\newblock \emph{\bibinfo{journal}{Nano Letters}} \textbf{\bibinfo{volume}{22}}\textbf{, }\bibinfo{pages}{1233--1241} (\bibinfo{year}{2022}).

\bibitem{48gu2023multi}
\bibinfo{author}{Gu, P.}, \bibinfo{author}{Wang, C.}, \bibinfo{author}{Su, D.} \emph{et~al.}
\newblock \bibinfo{title}{Multi-state data storage in a two-dimensional stripy antiferromagnet implemented by magnetoelectric effect}.
\newblock \emph{\bibinfo{journal}{Nature Communications}} \textbf{\bibinfo{volume}{14}}\textbf{, }\bibinfo{pages}{3221} (\bibinfo{year}{2023}).

\bibitem{37zhang2023spin}
\bibinfo{author}{Zhang, M.}, \bibinfo{author}{Hu, Q.}, \bibinfo{author}{Huang, Y.} \emph{et~al.}
\newblock \bibinfo{title}{Spin-lattice coupled metamagnetism in frustrated van der waals magnet {CrOCl}}.
\newblock \emph{\bibinfo{journal}{Small}} \textbf{\bibinfo{volume}{19}}\textbf{, }\bibinfo{pages}{2300964} (\bibinfo{year}{2023}).

\bibitem{35qing2020magnetism}
\bibinfo{author}{Qing, X.}, \bibinfo{author}{Li, H.}, \bibinfo{author}{Zhong, C.} \emph{et~al.}
\newblock \bibinfo{title}{Magnetism and spin exchange coupling in strained monolayer {CrOCl}}.
\newblock \emph{\bibinfo{journal}{Physical Chemistry Chemical Physics}} \textbf{\bibinfo{volume}{22}}\textbf{, }\bibinfo{pages}{17255--17262} (\bibinfo{year}{2020}).

\bibitem{36angelkort2009observation}
\bibinfo{author}{Angelkort, J.}, \bibinfo{author}{W{\"o}lfel, A.}, \bibinfo{author}{Sch{\"o}nleber, A.} \emph{et~al.}
\newblock \bibinfo{title}{Observation of strong magnetoelastic coupling in a first-order phase transition of {CrOCl}}.
\newblock \emph{\bibinfo{journal}{Physical Review B—Condensed Matter and Materials Physics}} \textbf{\bibinfo{volume}{80}}\textbf{, }\bibinfo{pages}{144416} (\bibinfo{year}{2009}).

\bibitem{38schaller2023pressure}
\bibinfo{author}{Schaller, A.~M.}, \bibinfo{author}{Bykov, M.}, \bibinfo{author}{Bykova, E.} \emph{et~al.}
\newblock \bibinfo{title}{Pressure-dependent distortions in quasi-two-dimensional magnetic {CrOCl} at low temperatures}.
\newblock \emph{\bibinfo{journal}{Physical Review B}} \textbf{\bibinfo{volume}{108}}\textbf{, }\bibinfo{pages}{104108} (\bibinfo{year}{2023}).

\bibitem{40zhang2022tuning}
\bibinfo{author}{Zhang, T.}, \bibinfo{author}{Zhang, Y.}, \bibinfo{author}{Huang, M.} \emph{et~al.}
\newblock \bibinfo{title}{Tuning the exchange bias effect in {2D} van der waals ferro-/antiferromagnetic {Fe$_{3}$GeTe$_{2}$/CrOCl} heterostructures}.
\newblock \emph{\bibinfo{journal}{Advanced Science}} \textbf{\bibinfo{volume}{9}}\textbf{, }\bibinfo{pages}{2105483} (\bibinfo{year}{2022}).

\bibitem{42yang2023unconventional}
\bibinfo{author}{Yang, K.}, \bibinfo{author}{Gao, X.}, \bibinfo{author}{Wang, Y.} \emph{et~al.}
\newblock \bibinfo{title}{Unconventional correlated insulator in {CrOCl}-interfaced bernal bilayer graphene}.
\newblock \emph{\bibinfo{journal}{Nature Communications}} \textbf{\bibinfo{volume}{14}}\textbf{, }\bibinfo{pages}{2136} (\bibinfo{year}{2023}).

\bibitem{33guo2024van}
\bibinfo{author}{Guo, Y.}, \bibinfo{author}{Li, J.}, \bibinfo{author}{Zhan, X.} \emph{et~al.}
\newblock \bibinfo{title}{Van der waals polarity-engineered {3D} integration of {2D} complementary logic}.
\newblock \emph{\bibinfo{journal}{Nature}} \bibinfo{pages}{1--7} (\bibinfo{year}{2024}).

\bibitem{25_49}
\bibinfo{author}{Forsberg, H.}, \bibinfo{author}{Songstad, J.}, \bibinfo{author}{Viljanto, J.} \emph{et~al.}
\newblock \href{https://api.semanticscholar.org/CorpusID:97887033}{\bibinfo{title}{On the structure of {CrOCl}}}.
\newblock \emph{\bibinfo{journal}{Acta Chemica Scandinavica}} \textbf{\bibinfo{volume}{16}}\textbf{, }\bibinfo{pages}{777--777} (\bibinfo{year}{1962}).

\bibitem{26_50}
\bibinfo{author}{Christensen, A.~N.}, \bibinfo{author}{Johansson, T.} \& \bibinfo{author}{Qu{\'e}zel, S.}
\newblock \href{https://api.semanticscholar.org/CorpusID:97548692}{\bibinfo{title}{Preparation and magnetic properties of crocl}}.
\newblock \emph{\bibinfo{journal}{Acta Chemica Scandinavica}} \textbf{\bibinfo{volume}{6}}\textbf{, }\bibinfo{pages}{1171--1174} (\bibinfo{year}{1975}).

\bibitem{27_51}
\bibinfo{author}{Coïc, L.}, \bibinfo{author}{Spiesser, M.}, \bibinfo{author}{Palvadeau, P.} \emph{et~al.}
\newblock \href{https://www.sciencedirect.com/science/article/pii/0025540881900866}{\bibinfo{title}{Chromium (iii) oxyhalides : Magnetic and optical properties. lithium intercalation}}.
\newblock \emph{\bibinfo{journal}{Materials Research Bulletin}} \textbf{\bibinfo{volume}{16}}\textbf{, }\bibinfo{pages}{229--236} (\bibinfo{year}{1981}).

\bibitem{30_54}
\bibinfo{author}{Angelkort, J.}, \bibinfo{author}{W\"olfel, A.}, \bibinfo{author}{Sch\"onleber, A.} \emph{et~al.}
\newblock \href{https://link.aps.org/doi/10.1103/PhysRevB.80.144416}{\bibinfo{title}{Observation of strong magnetoelastic coupling in a first-order phase transition of {CrOCl}}}.
\newblock \emph{\bibinfo{journal}{Physical Review B}} \textbf{\bibinfo{volume}{80}}\textbf{, }\bibinfo{pages}{144416} (\bibinfo{year}{2009}).

\bibitem{31_55}
\bibinfo{author}{Lenzlinger, M.} \& \bibinfo{author}{Snow, E.}
\newblock \bibinfo{title}{Fowler-nordheim tunneling into thermally grown {SiO$_{2}$}}.
\newblock \emph{\bibinfo{journal}{IEEE Transactions on Electron Devices}} \textbf{\bibinfo{volume}{15}}\textbf{, }\bibinfo{pages}{686--686} (\bibinfo{year}{1968}).

\bibitem{32_56}
\bibinfo{author}{Simmons, J.~G.}
\newblock \href{https://doi.org/10.1063/1.1702682}{\bibinfo{title}{{Generalized Formula for the Electric Tunnel Effect between Similar Electrodes Separated by a Thin Insulating Film}}}.
\newblock \emph{\bibinfo{journal}{Journal of Applied Physics}} \textbf{\bibinfo{volume}{34}}\textbf{, }\bibinfo{pages}{1793--1803} (\bibinfo{year}{1963}).

\bibitem{33_57}
\bibinfo{author}{Wang, Z.}, \bibinfo{author}{Guti{\'e}rrez-Lezama, I.}, \bibinfo{author}{Ubrig, N.} \emph{et~al.}
\newblock \bibinfo{title}{Very large tunneling magnetoresistance in layered magnetic semiconductor {CrI$_{3}$}}.
\newblock \emph{\bibinfo{journal}{Nature Communications}} \textbf{\bibinfo{volume}{9}}\textbf{, }\bibinfo{pages}{2516} (\bibinfo{year}{2018}).

\bibitem{58yao2023multiple}
\bibinfo{author}{Yao, F.}, \bibinfo{author}{Multian, V.}, \bibinfo{author}{Wang, Z.} \emph{et~al.}
\newblock \bibinfo{title}{Multiple antiferromagnetic phases and magnetic anisotropy in exfoliated {CrBr$_{3}$} multilayers}.
\newblock \emph{\bibinfo{journal}{Nature Communications}} \textbf{\bibinfo{volume}{14}}\textbf{, }\bibinfo{pages}{4969} (\bibinfo{year}{2023}).

\bibitem{38_59}
\bibinfo{author}{Wang, Z.}, \bibinfo{author}{Gibertini, M.}, \bibinfo{author}{Dumcenco, D.} \emph{et~al.}
\newblock \bibinfo{title}{Determining the phase diagram of atomically thin layered antiferromagnet {CrCl$_{3}$}}.
\newblock \emph{\bibinfo{journal}{Nature Nanotechnology}} \textbf{\bibinfo{volume}{14}}\textbf{, }\bibinfo{pages}{1116--1122} (\bibinfo{year}{2019}).

\bibitem{60telford2020layered}
\bibinfo{author}{Telford, E.~J.}, \bibinfo{author}{Dismukes, A.~H.}, \bibinfo{author}{Lee, K.} \emph{et~al.}
\newblock \bibinfo{title}{Layered antiferromagnetism induces large negative magnetoresistance in the van der waals semiconductor {CrSBr}}.
\newblock \emph{\bibinfo{journal}{Advanced Materials}} \textbf{\bibinfo{volume}{32}}\textbf{, }\bibinfo{pages}{2003240} (\bibinfo{year}{2020}).

\bibitem{61}
\bibinfo{author}{Long, G.}, \bibinfo{author}{Henck, H.}, \bibinfo{author}{Gibertini, M.} \emph{et~al.}
\newblock \href{https://doi.org/10.1021/acs.nanolett.9b05165}{\bibinfo{title}{Persistence of magnetism in atomically thin {MnPS$_{3}$} crystals}}.
\newblock \emph{\bibinfo{journal}{Nano Letters}} \textbf{\bibinfo{volume}{20}}\textbf{, }\bibinfo{pages}{2452--2459} (\bibinfo{year}{2020}).

\end{thebibliography}
\end{document}